\documentclass[a4paper]{aa}
\usepackage{graphicx}
\def\spose#1{\hbox to 0pt{#1\hss}}
\def\lta{\mathrel{\spose{\lower 3pt\hbox{$\mathchar"218$}}
     \raise 2.0pt\hbox{$\mathchar"13C$}}}
\def\gta{\mathrel{\spose{\lower 3pt\hbox{$\mathchar"218$}}
     \raise 2.0pt\hbox{$\mathchar"13E$}}}

\newcommand{\etal}{et al. }

\begin{document}

   \thesaurus{ 
        10.07.2;
        10.15.1;
        11.09.2;
        11.19.3;
        11.19.4;
        }
   \title{The Mass Function of Young Star Clusters in The Antennae}

   \author{U. Fritze -- v. Alvensleben}

   \institute{Universit\"atssternwarte G\"ottingen\\
              Geismarlandstr. 11, 37083 G\"otingen, Germany\\
              email: ufritze@uni-sw.gwdg.de
             }
   \authorrunning{Fritze -- v. Alvensleben}
   \titlerunning{Mass Function of YSCs in Antennae}

\date{Submitted Nov. 26, 1998}

\maketitle

\begin{abstract}
The Antennae is a pair of late type spirals in the course of merging. 
The interaction triggered an ongoing strong burst of star formation that also 
produced a large 
number of Young Star Clusters (YSCs), many of which seem to be young Globular Clusters 
(GCs). The observed Luminosity Function of the YSC system is a power-law. 
We use evolutionary synthesis models for star clusters in comparison with HST WFPC 
observations of the YSC system in the Antennae to analyse the mass function of this 
${\rm \sim 2 \cdot 10^8}$ yr young star cluster system. Properly accounting for age spread 
effects by individually age-dating the YSCs we find that the intrinsic Mass Function (MF) 
is log-normal in shape and in intriguing agreement with the MF of old GC systems. We 
discuss this MF in the context of cluster formation and dynamical effects 
in the tidal field of the parent galaxy and speculate about its future 
evolution.
\end{abstract}

\section{Introduction}
The Antennae (NGC 4038/39 = Arp 244) is a pair of late type spirals in the 
course of merging. It is the youngest in Toomre's (1977) 
dynamical age sequence of 11 interacting and merging systems and comfortably nearby. 
This makes it a 
well-studied system all over the wavelength range from X-ray through 
radio wavelengths.  The interaction has triggered a strong burst of star formation that 
has also produced a large 
number of $\sim 700$ bright star clusters as first observed with HST by Whitmore \& Schweizer 
(1995) ({\bf WS95}). Many of those bright clusters seem to be young Globular Clusters 
({\bf GCs}) due to their small effective radii and high luminosities. The Luminosity Function 
({\bf LF}) of this bright star cluster system looks like a power law 
${\rm \Phi(L) \sim L^{-1.78 \pm 0.05}}$ with no hint of a turnover at fainter magnitudes 
down to the completeness limit of ${\rm V = 22.3}$ mag which, at the distance of The 
Antennae (19.2 Mpc  for H$_0 = 75$), corresponds to ${\rm M_V = -9.6}$ mag (WS95). 
In Fritze -- v. Alvensleben (1998) (hereafter {\bf Pap.I}) I analysed WS95's HST data 
with evolutionary synthesis models for single burst single metallicity 
populations ({\bf SSPs}) using stellar evolutionary tracks for various metallicities 
from the Geneva group and a Scalo IMF from 0.15 to 60 M$_{\odot}$ (Fritze -- v. Alvensleben 
\& Burkert 1995, {\bf FB95}). In analogy to 
the YSC system in NGC 7252 which, like the Antennae, is an Sc -- Sc merger, we assume 
that the metallicity of the YSC population that forms out of the spirals' ISM is ${\rm \sim 
\frac{1}{2} \cdot Z_{\odot}}$ (Fritze -- v. Alvensleben \& Gerhard 1994). 
In Pap.I individual ages are determined from (${\rm V - I}$) colours for the 550 
star clusters with I -- band detections. The age distribution clearly reveals 
two populations of clusters, 
$\sim 70$ old GCs from the original spirals, and $\sim 480$ {\bf Young} Star Clusters 
({\bf YSCs}) with ages in the range $(0 - 4) \cdot 10^8$ yr formed in the ongoing 
interaction-induced starburst. Only the secondary population of YSCs will be 
considered in the following. 
Meurer (1995) was the first to point out the possible importance of age spread effects 
on the future evolution of the LF of a YSC system. With individual YSC ages and the 
fading in various passbands as given by our SSP models 
we were able to calculate the future evolution over a Hubble time of the LF and of the 
colour distribution of the YSC system under the unrealistic assumption that all clusters will 
survive. At an age of $\sim 12$ Gyr, when age differences of the order of $10^8$ yr do no longer 
play any role, the LF is shown to be undistinguishable from a typical GC LF (i.e. Gaussian with a 
turnover at ${\rm M_V \sim -6.9}$ mag -- appropriate for the metallicity [Fe/H] $\sim -0.3$ 
(Ashman \etal 1995) -- and ${\rm \sigma(M_V) = 1.3}$ mag). The colour distribution will 
by then also be compatible with the one observed on old GC systems. 
While accounting for stellar mass loss, this modelling, however, did 
not take into account dynamical effects like two-body relaxation, dynamical friction, 
or disk shocking that might 
act heavily on a YSC system over a Hubble time. 
Already for the Milky Way, where the potential is rather well known, it is difficult and not 
uncontroversial to model the dynamical processes on individual clusters 
(e.g. Chernoff \& Weinberg 1990, Fukushige \& Heggie 1995, Gnedin \& Ostriker 1997, 
Vesperini 1997). This is, 
of course, even more difficult, if not impossible, in an ongoing merger like The Antennae. 
Before one can try and quantify the efficiency of various destruction mechanisms, the 
intrinsic MF underlying the presently observed LF has to be determined. 
Knowing individual cluster ages offers the possibility to use M/L values from 
evolutionary synthesis models to derive the Mass Function ({\bf MF}) underlying the presently observed 
LF and this is what we attempt in this Letter. 
Since the age distribution of the YSCs is strongly peaked within $\leq 2 \cdot 10^8$ yr and the 
YSCs have a mean distance to the galaxy center of $\sim 3.5$ kpc, a YSC on average cannot have 
had more than 1 or 2 revolutions. We do not expect the MF therefore to already 
be significantly affected by cluster destruction processes. Rather we expect the presently 
derived MF to reflect the MF produced by the cluster formation process. 

The mass spectra of molecular clouds, molecular cloud cores, open clusters, 
and the LF of giant HII regions (in non-interacting galaxies) all are  
power laws with exponents $\alpha$ in the range 
$\alpha \sim -1.5 \ \dots \ -1.7$ (e.g. Solomon \etal 1987, Lada \etal 1991, 
Kennicutt 1989, 
see Harris \& Pudritz 1994 and Elmegreen \& Efremov 1997 for overviews) 
as is the mass spectrum of open clusters in the 
Milky Way and the LMC (e.g. van den Bergh \& Lafontaine 1984, Elson \& Fall 1985). 
Both the LF and the MF of old GC systems are Gaussians with typical 
parameters ${\rm \langle M_V \rangle \sim -7.3}$ mag, $\sigma \sim 1.3$ mag, 
and ${\rm \langle Log~(M/M_{\odot}) \rangle \sim 5.5}$, ${\rm \sigma \sim 0.5}$, 
respectively (e.g. Ashman \etal 1995).   

The question immediately arises: Is the 
transition from a power law molecular cloud mass spectrum to a Gaussian old GC mass spectrum 
performed in the star/cluster formation process or by secular destruction effects 
within a GC system ? Or, else, is already the mass spectrum of molecular clouds 
or molecular cloud cores different in strongly interacting and starbursting galaxies 
from what it is in normal spirals ?

\section{The Mass Function of the YSCs in The Antennae}
On the basis of individual YSC ages we use our SSP models giving ${\rm M/L_{\lambda}}$ 
in the passbands $\lambda = $ UBVRIK as a function of time to derive masses of individual clusters  
from their observed V -- luminosities. 

This is done for all the 393 YSCs with ages $\leq 4 \cdot 10^8$ yr and V -- luminosities 
brighter than the completeness limit ${\rm M_V = -9.6}$ mag. It is stressed that our model 
${\rm M/L}$ - values include the decrease in cluster mass due to stellar mass loss 
(cf. Pap.I), but not that due to the evaporation of stars 
from the cluster. The MF we recover 
in this way from the presently observed LF is presented in Fig.1. A Gaussian with 
${\rm \langle Log~(M_{YSC}/M_{\odot}) \rangle = 5.6}$ and ${\rm \sigma = 0.46}$, normalised to the 
number of YSCs in the histogram, is overplotted. The intrinsic MF we obtain for the YSCs 
brighter than the completeness limit in The Antennae clearly looks log-normal in shape 
with the maximum at a mean YSC mass of ${\rm \sim 4 \cdot 10^5~M_{\odot}}$. Stellar mass loss 
within the clusters from the present mean age of $\sim 2 \cdot 10^8$ 
yr through an age of $\sim 12$ Gyr will lead to a decrease in mass of $\lta 15 ~\%$ 
for a Scalo -- IMF ($< 10 \%$ for Salpeter).

\begin{figure}
\includegraphics[width=\columnwidth]{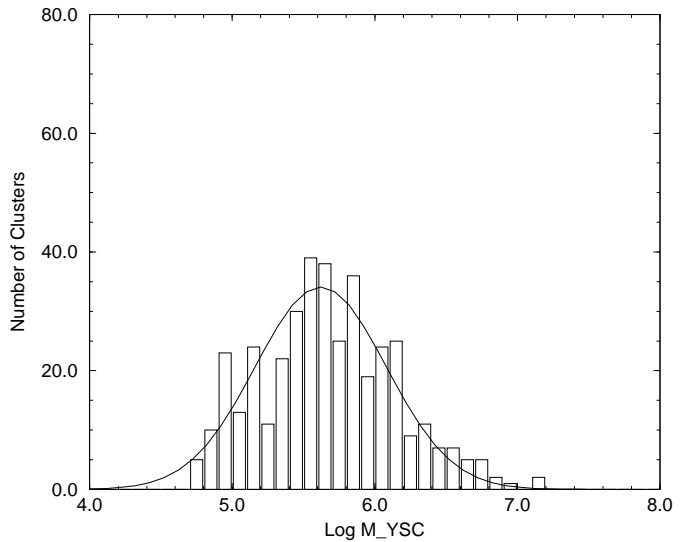}
\caption{Mass Distribution of the YSCs in The Antennae: 393 YSCs brighter than the completeness 
limit. A Gaussian with ${\rm \langle Log~(M_{YSC}/M_{\odot}) \rangle = 5.6}$ and 
${\rm \sigma = 0.46}$ is overplotted, normalised to the 
number of YSCs in the histogram.}
\end{figure}

Thus, without any destruction or evaporation effects the mean mass of the secondary GC system 
in The Antennae at the age of a Hubble time would be ${\rm \sim 3.4 \cdot 10^5 
~M_{\odot}}$. 
A cluster with this mean mass would have ${\rm M_V = -6.9}$ mag at an age of 12 Gyr according to 
our models. This is the position of the maximum of the Gaussian YSC LF we obtain at a 
hypothetic YSC age of 12 Gyr (cf. Fig. 6a in Pap.I). The agreement is no surprise since 
this is, in fact, the way how we obtained the parameters for the Gaussian in Fig.1. We stress 
that these parameters are derived from the LF in Pap.I and are not the result 
of any fit to the cluster MF in Fig.1.

Remarkably enough, the parameters of this Gaussian -- which in Fig.1 is seen to reasonably describe 
the MF of the secondary GC population -- are quite similar to those given by Ashman \etal (1995) for the 
Milky Way 
and M31 GC systems. Using evolutionary synthesis results from Worthey (1994) for ${\rm M/L_V}$, 
Ashman \etal find ${\rm \langle Log~(M/M_{\odot}) \rangle = 5.47}$ and ${\rm \sigma = 0.50}$ 
for the Milky Way and ${\rm \langle Log~(M/M_{\odot}) \rangle = 5.53}$ and ${\rm \sigma = 0.43}$ 
for M31. 
The MF in Fig. 1 thus seems compatible with the bulk of the YSCs really being young GCs 
rather than open clusters or associations, as was already indicated by their small 
effective radii and high luminosities (WS95). 

From the model side, 
uncertainties in the determination of YSC ages (and hence masses) 
on the basis of their ${\rm (V - I)}$ colours 
are dominated by uncertainties in the YSC metallicities. 
Age uncertainties due to metallicity uncertainty 
(${\rm \frac{1}{2} \cdot Z_{\odot} \lta Z_{YSC} \lta Z_{\odot}}$) are estimated to be 
of the order of $\pm 15~\%$. The uncertainty in M/L at ${\rm Z = \frac{1}{2} \cdot Z_{\odot}}$ 
due to the age uncertainty is $\sim 8$ \% and the uncertainty in M/L at all ages 
$\lta 4 \cdot 10^8$ yr due to the metallicity uncertainty is $\lta 5$ \%. 
This leads to an overall uncertainty in the M/L of $\sim 10$ \%. 

Measurement uncertainties are $\sim \pm 0.2$ mag for the observed ${\rm (V - I)}$ colours 
and $\lta 0.15$ 
mag for V magnitudes for YSCs brighter than the completeness limit. 
The uncertainties that the inhomogeneous reddening over the body of NGC 4038/39 
brings along for the derived cluster ages and masses are difficult to quantify. 
Only a global average value is given for the internal reddening of the YSC population 
in NGC 4038/39 (${\rm E_{V-I} \sim 0.3}$ mag (WS95)) and 
applied before our age-dating, but dust lanes 
and unrelaxed structures are seen all over NGC 4038/39. 
The very good agreement ($\leq 10^7$ yr) between 
ages determined from ${\rm (U - V)}$ and ${\rm (V - I)}$ colours, however, indicates that 
for the bright clusters seen on the short U exposure the reddening does {\bf not} seem to 
significantly deviate from the average reddening we use. 

It should be noted in this context that inclusion of YSCs fainter than the completeness limit -- 
which tend to have significantly larger observational uncertainties -- 
does {\bf not} affect the agreement with the Gaussian in Fig.1 or its parameters (beyond 
the normalisation to the number of clusters in the histogram).

{\bf We conclude that the secondary GC population in The Antennae is formed with 
a log-normal mass distribution very similar to the one in the Milky Way or M31. 
It is not necessarily the secular evolution but rather 
the cluster formation process that produces the Gaussian mass spectrum observed 
on old GC systems.}

Since uncertainties in the MF from observational errors cannot be calculated in any 
straightforward way in the analysis presented here, we are currently trying an 
independent approach. We draw YSCs at random from different intrinsic MFs (Gaussians, power-laws), 
randomly assign ages to them from different age distributions (clusters formed uniformly over 
the burst duration or at an increasing or decreasing rate), and calculate their present 
luminosities and colours to which, then, observational errors can be added, 
again at random from the 
observed luminosity-dependent error distribution. Comparison of the resulting model LFs and 
colour distributions with 
the observed ones should then allow to constrain the intrinsic MF (Kurth \etal, {\sl in prep.}). 
A somewhat similar procedure is used for YSCs in NGC 7252 and NGC 3921 by 
Miller \& Fall (1997), who find that power-laws are preferred over Gaussians for the 
MFs of these YSC systems.

\section{Comparison with YSCs in NGC 7252 and NGC 3921}
NGC 7252 and NGC 3921 are the two oldest merger remnants from Toomre's (1977) list. Large enough 
YSC systems have been detected in both of them to define the bright end of their LFs 
(Whitmore \etal 1993, 
Miller \etal 1997, Schweizer \etal 1996). Distances, however, are larger than to NGC 4038/39 
by factors 3 and 4 for NGC 7252 and NGC 3921, respectively, pushing the completeness limit to 
significantly higher luminosities. 
The higher mean ages of the YSCs in these cases ($650 - 750$ Myr for 
NGC 7252 and $250 - 750$ Myr 
for NGC 3921, depending on metallicity)
add another argument in favour of them being young GCs, since 
they have survived ${\rm \gg 10 \cdot t_{cross}}$ 
(cf. Schweizer \etal 1996). 

The LF we {\bf calculate} for the YSCs in The Antennae at an age of $\sim 1$ Gyr does show 
a marginal turnover at the expected ${\rm \langle M_V \rangle} \sim -9.9$ mag 
for YSCs brighter than the completeness limit, indicating that by this age the 
distortion of the LF with respect to the MF due to age spread 
effects from the finite YSC formation timespan (of the 
order $200 - 400$ Myr) is already less important. 

In order to estimate if a turn-over in their YSC LFs could be expected in NGC 7252 and 
NGC 3921, 
we assume that their YSC systems have the same MF as the YSC system in The Antennae. Then, 
we can calculate the mean absolute magnitude ${\rm \langle M_V \rangle}$ from 
${\rm \langle M_{YSC} \rangle}$ at 
the above-quoted mean ages. We obtain  
${\rm \langle M_V \rangle} = -10 \ \dots \ -9.5$ mag and 
${\rm \langle M_V \rangle} = -9.5 \ \dots \ -9$ mag for YSCs in NGC 7252 and in NGC 3921, 
respectively. These luminosities 
are close to the 90 \% completeness limiting magnitudes of 
$-9.5$ (PC) and $-8.5$ (WF) for NGC 7252 (Miller \etal 1997) and of $-9.0$ for 
NGC 3921 (Schweizer \etal 1996). 
Together with the difficulty of disentangling the old GC population 
from the YSCs at these advanced ages the fact that no turnover is detected in the LFs for NGC 7252 
and NGC 3921 does not seem to rule out a Gaussian MF similar to the one we obtain in 
The Antennae.

\section{Dynamical Evolution of the YSC System}
Quantitatively, nothing is known about the external dynamical effects on GC systems in 
interacting and merging galaxies. For the Milky Way, where dynamical modelling on the basis of 
the Galactic potential is possible and has been done by many groups (e.g. Chernoff \& Weinberg 
1990, Gnedin \& Ostriker 1997, Fukushige \& Heggie 1995, Vesperini 
\& Heggie 1997, Vesperini 1997, 1998) it is clear that the GC system 
observed today is only ``the hardiest survivors of a larger original population'' (Harris 1991). 
Vesperini \& Heggie (1997) present N-body simulations including 
effects of stellar evolution, 
two-body relaxation, disk shocking, and the tidal field of  
the Milky Way.  
Studying the secular evolution of a number of GC systems with different initial MFs, 
Vesperini (1998) shows that {\bf if the initial MF is log-normal then this log-normal shape 
and its parameters 
are conserved over a Hubble time} despite the destruction of a large part ($\sim 50~\%$) 
of the original GC population. 
While evaporation and dynamical friction preferentially destroy low and high mass clusters, 
respectively, both processes balance each other in the case of a log-normal 
($=$ equilibrium) initial GC MF, so that 
no net selective destruction of specific GC masses results.  
If the GC destruction processes were similar in the Antennae and in 
the Milky Way, and if, as indicated in Fig.1, the initial GC MF really were close to Vesperini's 
equilibrium MF, then the Gaussian MF could survive a Hubble time with its parameters 
${\rm \langle Log~(M/M_{\odot}) \rangle}$ and ${\rm \sigma}$ essentially unchanged  despite 
the likely destruction of a large fraction of the YSC system seen today. 

It will be interesting to analyse more YSC systems in the way we did in order to see if and in 
how far the YSC MF is universal or depends on parameters of the progenitor galaxies or the 
interaction event. Moreover, studying secondary GC systems on an age sequence of interacting, 
merging and merger remnant galaxies should allow to directly ``observe'' both the time evolution 
of the LF and of the MF and thus offer a unique possibility to study the 
effects of dynamical processes {\sl in situ}. The turnover of an old GC population at ${\rm 
M_V \sim -7.2}$ mag occurs around ${\rm V \sim 24.5}$ at Virgo cluster distances and is well 
within the reach of 10m telescopes. HST imaging allows to identify clusters in 
regions not too crowded for ground-based MOS.

\section{Discussion and Open Questions}
The uncertainties in the metallicities and -- more important -- in the individual reddening of 
YSCs in The Antennae can be substantially reduced with spectroscopic observations of 
YSC selected from the HST images. With MOS facilities on 10m class telescopes this should 
be possible in the nearest future. 

One question that poses itself immediately in the context of the cluster formation 
mechanism is whether the molecular cloud mass spectrum in a strongly interacting and starbursting 
galaxy like The Antennae does really have the same power-law form as that in an undisturbed spiral 
forming stars at a level orders of magnitude lower? 
Elmegreen \& Efremov (1997) pointed out that the high pressure of the ambient ISM produced 
in a strong interaction might favor the production of more massive clouds. 
Information about the molecular cloud mass spectrum in merger-induced strong starbursting 
systems seems a prerequisite for the study of the star and cluster formation processes.

First spectroscopic observations of a handful of YSCs in NGC 7252 (Schweizer \& Seitzer 1993, 
1998) confirms the metallicity of ${\rm (\frac{1}{2} - 1) \cdot Z_{\odot}}$ (cf. FB95) 
we predicted on the basis of the progenitor galaxies' ISM abundances. This enhanced 
metallicity (with 
respect to the primary GC population) raises the question in how far the secondary 
GC formation process is comparable to the primary one in the early Universe? 

\section{Preliminary Conclusions}

{\bf 1.} The MF of the YSCs in The Antennae seems to be log-normal with parameters 
${\rm \langle Log~(M/M_{\odot}) \rangle}$ and ${\rm \sigma}$ very similar to those 
of the Milky Way GC system. \hfill\break
{\bf 2.} This suggests that the cluster formation process and not the dynamical evolution produce 
the Gaussian MF. \hfill\break
{\bf 3.} The close agreement 
of the YSC MF we obtain with Vesperini's equilibrium initial GC MF seems to indicate that 
the shape and parameters of this MF may survive a Hubble time despite destruction of 
a large fraction of today's YSCs. \hfill\break
{\bf 4.} In the older merger remnants NGC 7252 and NGC 3921 the completeness limit is 
close to the turn-over luminosity expected in case their MFs were similar to 
the one in The Antennae. 

As long as the impact of observational colour and luminosity uncertainties on the MF we derive 
cannot be quantified rigorously, our conclusions have to remain preliminary.

\begin{acknowledgements}
I gratefully acknowledge a very prompt and constructive report from an anonymous referee 
and I'm deeply indebted to Bruce Elmegreen for critical and stimulating discussions. 
\end{acknowledgements}

\end{document}